\documentclass[twocolumn]{aastex63}

\usepackage{epstopdf}
\usepackage{amsmath}

\shorttitle{anionic cyanopolyynes}
\shortauthors{ Gonzalez-Sanchez and Gianturco}

\graphicspath{{./}{figures/}}

\begin{document}

\author{L. Gonz\'{a}lez-S\'{a}nchez}
\affiliation{Departamento de Qu\'{i}mica F\'{i}sica, University of Salamanca, Plaza de los Ca\'{i}dos sn, 37008, Salamanca, Spain}
  
 \author{A.Veselinova}
\affiliation{Departamento de Qu\'{i}mica F\'{i}sica, University of Salamanca, Plaza de los Ca\'{i}dos sn, 37008, Salamanca, Spain}
  
 \author{A. Mart\'{i}n Santa Dar\'{i}a}
\affiliation{Departamento de Qu\'{i}mica F\'{i}sica, University of Salamanca, Plaza de los Ca\'{i}dos sn, 37008, Salamanca, Spain}

 \author{E. Yurtsever}
\affiliation{Department of Chemistry, Koc University Rumelifeneriyolu, Sariyer TR 34450, Istanbul, Turkey}

\author{R.Biswas}
\affiliation{School of Chemical Sciences, National
Institute of Science Education and Research (NISER) Bhubaneswar}

\author{K.Giri}
\affiliation{Department of Computational Sciences, Central University of Punjab, Bathinda, Punjab 151401, India}

\author{N. Sathyamurthy}
\affiliation{Indian Institute of Science Education and Research Mohali, SAS Nagar, Punjab 140306, India}

\author{U. Lourderaj}
\affiliation{School of Chemical Sciences, National
Institute of Science Education and Research (NISER) Bhubaneswar}

\author{R. Wester}
\affiliation{Institut f\"{u}r Ionenphysik und Angewandte Physik, Universit\"{a}t Innsbruck Technikerstr. 25 A-6020, Innsbruck, Austria}

 \author{F. A. Gianturco}
\affiliation{Institut f\"{u}r Ionenphysik und Angewandte Physik, Universit\"{a}t Innsbruck Technikerstr. 25 A-6020, Innsbruck, Austria}
\email{francesco.gianturco@uibk.ac.at}

\title{Computed Rotational Collision Rate Coefficients  for Recently Detected Anionic Cyanopolyynes }

\date{\today}

\begin{abstract}
We report new results from quantum calculations  of energy-transfer processes taking place in interstellar environments and involving two newly observed molecular  species:  C$_5$N$^-$ and C$_7$N$^-$ in collision with He atoms and the p-H$_2$ molecules. These species are part of the anionic molecular chains labeled as cyanopolyynes which have  been observed over the years in  molecule-rich Circumstellar Envelopes (CSEs) and in molecular clouds. In the present work, we first carry out  new $ab$ $initio$ calculations for the C$_7$N$^-$  interaction potential with He atom and  then obtain  state-to-state rotationally inelastic cross sections and rate coefficients involving the same transitions which have been observed experimentally by emission in the interstellar medium (ISM) from both of these  linear species. For the C$_5$N$^-$/He system  we extend the calculations already published in our earlier work (see reference below) to compare more directly the two molecular anions. We  extend further the quantum  calculations by also computing in this work  collision rate coefficients  for the hydrogen molecule interacting with C$_5$N$^-$, using our previously computed interaction potential. Additionally, we obtain the same rate coefficients for the C$_7$N$^-$/H$_2$  system by using a scaling procedure that makes use of  the new C$_7$N$^-$/He rate coefficients, as discussed in detail in the present paper. Their significance in affecting internal state populations in ISM environments where the title anions have been found is analyzed by  using the concept of critical density indicators. Finally, similarities and differences between such species and the comparative efficiency of their collision rate coefficients are discussed. These new calculations suggest that, at least for the case of these longer chains, the rotational populations could reach local thermal equilibrium (LTE) conditions within their observational environments.
\end{abstract}



\section{Introduction}
\label{sec:intro}

  The study of the molecular complexity of the interstellar medium (ISM) provides us with fundamental information on the chemical storage of species and on the energy available for the formation of stars and planets. It also gives us useful indication about the chemical inventory that the primitive Earth might have inherited and about the formation paths of many of the species increasingly discovered by direct observations. 
  
 In recent years, several linear C-bearing and (C,N)-bearing chains of  molecular anions have been detected at various sites in the  circumstellar envelopes (CSE) and in molecular clouds. Specifically: CN$^-$\cite{Agundez2010}, C$_3$N$^-$ \cite{Thaddeus2008}, C$_5$N$^-$ \cite{Cernicharo2008, Cernicharo2020}, C$_4$H$^-$ \cite{Cernicharo2007, Agundez2008}, C$_6$H$^-$ \cite{McCarthy2006}, and C$_8$H$^-$ \cite{Brunken2007, Remijan2007}. Those reported so far  constitute different terms of the general linear chains associated with the polyynes and cyanopolyynes species. Interestingly, very recently two additional (C,N)-bearing chains have been detected. They involve the two longest sequences of the linear-chain series  observed so far : C$_{10}$H$^-$  \cite{Rem2023} and C$_7$N$^-$ \cite{Cern2023}.

 The chemistry of formation of the cyanopolyyne chains has also been the object of several studies and speculations (e.g. see: \cite{Khamesian2016, Cernicharo2020}). The possible formation of anions  from the neutral radical via a Radiative Electron Attachment (REA) process has been considered in some detail (as quoted in: \cite{18JeGiWe.Cnm}) while a more direct chemical route by reaction of the HC$_5$N with H$^-$ has also been put forward by our group \cite{15SaGiCa.LM}. Another possible mechanism of formation could be the reaction between N atom and C$_n^-$, as discussed in:  \cite{Cernicharo2020}. For the present case of the C$_5$N$^-$ our calculated rates were found to be large enough to be relevant within the chemical evolution producing these anions. As also discussed by \cite{Cernicharo2020} the much larger rate constant for the REA formation path expected for C$_5$N compared to C$_3$N is the main reason why C$_5$N$^-$  has been estimated to be much more abundant with respect to the neutral than C$_3$N$^-$ (the abundance ratios of the C$_n$N /C$_n$N$^-$ are $\sim$ 140 and 2.3 for $n=3$ and $n=5$, respectively, in the Taurus molecular cloud 1 (TMC-1)). For the case of the longer chain of the C$_7$N$^-$ anion its abundance ratio with the next smaller chain discussed in the present work, the C$_5$N$^-$ anion, was found in the molecular cloud TMC-1 to be around 5 in  \cite{Cern2023}. On the other hand, the same C$_5$N$^-$/C$_7$N$^-$ ratio in the IRC -10216 was indicated in the same study to be around 2.4, i.e. not very different from the situation in the TMC-1 environment. In the final analysis, however, the various options put forward by the current literature about the chemical paths to the formation of the cyanopolyynes have not yet coalesced into a unique proof of the chemical  mechanism for the title anions or for their smaller partners in the series.
 
Astrophysical observation of these molecular anions relies heavily on the spectroscopic investigation of their properties in the laboratory and on matching sighted lines with those observed in the earthly experiments. Additionally, to carry out  astrophysical modeling of molecular population evolution from their distributions over a variety of internal states, important indicators are provided by the rate coefficients for the probability of rotational state-changes in molecular species induced by their interaction with He and H$_2$, both partners being present in substantial amounts in the observational environments of these anions. The collision-induced  occurrence of non-LTE (Local Thermal Equilibrium) population distribution for  rotational states in molecular partners could, in fact, be a significant path  to produce radiative emissions from different excited spectral lines in the observational microwave regions.  Measurement of these rate coefficients in the laboratory is still challenging, while  available quantum  methods can be used to model such processes  and then enter their results within networks of the kinetics of the underlying chemistry or of the internal-state populations in different environments. For example, Botschwina and Oswald \cite{Botschwina2008} had computationally determined the structural characteristics of some of these species to help observational sighting in the ISM or in the laboratory of specific emission lines

In the present work we focus more specifically on the collision-induced rotational state-changing processes associated with the observed emission lines involving the  C$_5$N$^-$/C$_7$N$^-$ anions, those discussed in the recent publications of \citep{Cernicharo2020, Cern2023} respectively. We thus compute the inelastic rate coefficients activated by interaction with He atoms and with p-H$_2$ molecules over the same range of transitions, and  further calculate the Einstein A-coefficients of spontaneous emissions between the same rotational  levels of these molecular anions. We then analyze the behaviour of the critical densities over the temperature range in the molecular clouds where the observations were made and discuss the consequences that such indicators suggest for the local distributions among rotational states in both molecules. The  state-changing collision rate coefficients are obtained using exact quantum computations  in the cases of  both the C$_5$N$^-$ and the C$_7$N$^-$ anions in collision with the He atoms. For the H$_2$ collision partner exact calculations are  carried out for the C$_5$N$^-$/H$_2$ system, while  a scaling procedure is implemented for the  H$_2$ collision rates with the C$_7$N$^-$ longer anion, making use of the new rates from the C$_7$N$^-$/He system and from the C$_5$N$^-$/H$_2$/He systems, as discussed in detail later in this paper. 
The details of the calculations and the analysis of the obtained results will be presented in the following Sections, while our conclusions will be given in the last Section.

\section{The Ab initio Computed Interaction Potential} 
\label{sec:pes}

The calculations  of the new Potential Enery Surface (PES) for the C$_7$N$^-$/He system were carried out using the GAUSSIAN09 set of codes as in \cite{Gauss09} with the UCCSD(T) approach based on initial UHF orbital expansion using the aug-cc-pVTZ for the carbon and nitrogen atoms, while for the He atomic partner the more extended basis set was of QZ-Vpp quality.For further details on the atomic basis we have employed see: \cite{89DuJr} and \cite{05WAhl} . The usual BSSE correction was applied: \cite{Gauss09}, \cite{94DeKnxx}, \cite{94WoDuxx}. No convergence problems were encountered in the calculations. In all the $ab$ $initio$ calculations we used an angular grid of 5$^{\circ}$ intervals. The optimized bond distances in the C$_5$N$^-$ partner were taken from our  previous calculations described in our earlier publication \cite{BGG23} where all the details of this anion's interaction with both He and H$_2$ were reported. Hence, we shall not be repeating this information in this Section, where we shall discuss instead the new calculations we have carried out for the longer anionic chain.

In the case of the C$_7$N$^-$ partner, the optimized bond distances were obtained from such new calculations. Hence, if one takes the carbon atom at the extreme left as the origin of the $x$-coordinate, then the C$_1$-C$_2$ distance was (all values in \AA\,): 1.2648, the C$_2$-C$_3$ distance: 1.4062; the C$_3$-C$_4$ bond: 1.2413; the C$_4$-C$_5$ bond: 1.2784; the C$_5$-C$_6$ bond: 1.2318; the C$_6$-C$_7$ bond: 1.3643; the N-C$_7$ distance: 1.1707 for a total length of the linear cyanopolyyne of 8.9575 and the c.o.m. (center-of-mass) location at 4.6002. The resulting interaction with the He atom was  described via  the distance $R$ from the c.o.m. and the polar angle $\theta$, both being  the usual  2D Jacobi coordinates  which describe the  present PES.

The radial range of the  raw points for the variable $R$ was taken from initial values of around 6.2 \AA\,  depending on the particular angular orientation, and extended out to   20.0 \AA\,. Radial intervals between 0.1 to 1.0 \AA\ were used depending on the distance and angle regions from the c.o.m. A variable number of radial values for each angular orientation were obtained as raw points, for a total range of 37 angular values. Note that the long, linear structure of the anionic species, and hence the location of the center-of-mass at the large distance mentioned earlier, are responsible for the interaction to become highly repulsive at fairly large distances for the atomic partner's approach to the c.o.m. We shall further show below that the actual radial range of the fitted potentials came closer to the c.o.m positions for some of the angles needed in the scattering calculations. It is important to note at this point that the \emph{ab initio} computed points are not symmetrically distributed along the distance coordinate due to the large anisotropy of the system, depending on the entrance angle of the He atom. This specific feature of the forces at play will be further discussed below.

This lack of points at short distances was therefore improved and completed by extrapolating with an exponential function for every angle up to 2~\r{A} with respect to the c.o.m. 
In order to generate a uniformly regular grid of points, spline functions were used on the \emph{ab initio} and on the extrapolated data sets to produce a total of 82 distance ($R$) points along the 37 angles ($\theta$), summing up to a total of 3034 points which were further fitted to generate the final PES. 
The intrinsic problem of having 'ghost points' at very short distances, even without them having physical sense because of their  being inside the anionic atoms' chain itself, is their extremely large energy values which accumulate over  short radial intervals. 
Such huge values cause mathematical instabilities when trying to fit them via an analytical function. 
To overcome this problem, we scaled the points by applying the simple procedure implemented by Faure \emph{et al.} \cite{faure2011potential} and already used in similar systems \cite{khadri2020low,sahnoun2018van,masso2014hco}.
It has the advantage that one can now reproduce the original \emph{ab initio} points in the lower energy range of the PES (the most relevant range for the scattering calculations) while one only scales the higher energy values that would not be sampled during the scattering events. Thus, one can write the following sequence, 
\begin{align}
&
   V_s = V_a  \quad \quad \quad \quad  \quad \quad \quad \quad  \quad \quad \quad \quad  \quad  \,\,
   \quad \quad \quad \text{for}   V_a \leq V_{T1} \nonumber \\
   &
   V_s = V_{T1} + (V_{T2}-V_{T1}) ~S \bigg( \frac{V_a-V_{T1}}{V_{T2}-V_{T1}} \bigg)
   \quad \quad \quad \text{for}   V_{T1} < V_a \leq V_{T2} \\
   &
   V_s = V_{T1} + \frac{2}{\pi} (V_{T2}-V_{T1}) \quad \quad \quad \quad \quad \,\,\,\,\,
   \quad \quad \quad \text{for}   V_a > V_{T2} \nonumber 
\end{align}
where $V_s$ is the scaled potential, $V_a$ is the \emph{ab initio} energy and $V_{T1}$ and $V_{T2}$ refer to two threshold energy values, which for this particular case are 2000~cm$^{-1}$ and 10000~cm$^{-1}$, respectively.
The  switching function is given by the $S$ function defined as 
\begin{align}
    S(x) = \frac{2}{\pi} \sin 
    \bigg[ 
    \frac{\pi}{2} \sin \bigg( \frac{\pi}{2} x \bigg) 
    \bigg]   \; ,
\end{align}
and it guarantees the proper transition between the three different scaling windows.

In the standard procedures employed for solving the Coupled-Channel (CC) scattering equations, it is usually convenient, as discussed below,  to expand the interaction potential $V(R,\theta)$ into orthogonal angular functions  \citep{60ArDaxx, 79Secrxx, 97KoHoffxx,79FaGxx}. Hence, for a  more direct, and quantitative evaluation  of  the spatial anisotropy around the C$_5$N$^-$  and the C$_7$N$^-$ linear anions it is useful to   represent the raw, 2D grid of points  from the  $ab$ $initio$ calculations in terms of the familiar Legendre polynomials in their standard ($R$,$\theta$) form. In the case of the new calculations discussed above, using the 3034 `scaled points' we were  able to expand the PES via the usual orthogonal Legendre polynomials for the $\theta$ angle:
\begin{equation}
   V_{\rm FIT} = V(R,\theta) = 
   \sum_{\lambda=0}^{\lambda_{\rm max}} 
   V_{\lambda}(R) P_{\lambda}(\cos\theta)
   \label{eq:PES_fit}
\end{equation}

The fitted potential is restricted to a finite range of $R$ values and therefore an extrapolation to an asymptotic  form is needed to describe the long-range interaction forces. 
The analytical form used for the long-range is
\begin{align}
    V_{\rm LR} (\theta) =
    - \frac{\alpha_0}{2R^4} + \frac{2\alpha_0 \mu}{R^5} \cos{\theta} \; ,
    \label{eq:PES_lr}
\end{align}
where $\alpha_0 = 1.41~a_0^3$, is the polarizability of the He atom and $\mu =7.5 D $, from \cite{Cern2023}, is the rigid rotor dipole moment of the C$_7$N$^-$ anionic partner.

To ensure the smooth transition between the $V_{\rm FIT}$ fitted PES (Eq.~\ref{eq:PES_fit}) and the $V_{\rm LR}$ long-range term (Eq.~\ref{eq:PES_lr}) in the construction of the $V_{\rm f}$ final PES, the switching procedure which was already  used for C$_5$N$^-$/He \cite{BGG23} has also  been implemented for the longer chain anion:
\begin{align}
    V_{\rm f} = 
    f_s  V_{\rm FIT} + (1-f_s) V_{\rm LR}
\end{align}
where the switching function is 
\begin{align}
    f_s (R) =
    \frac{1}{e^{\frac{(R-R_0)}{\Delta R} + 1}} \; ,
\end{align}
         with $R_0 = 18.5$ and $\Delta R = 1.2$ \r{A}.

A pictorial view of the spatial shape of the interaction forces for C$_7$N$^-$/He is reported in  Figure 1, where the energy isolines are depicted around the linear, rod-like structure of the anionic partner.

\begin{figure}[!ht]
   \centering
   \includegraphics[width=0.50\textwidth]{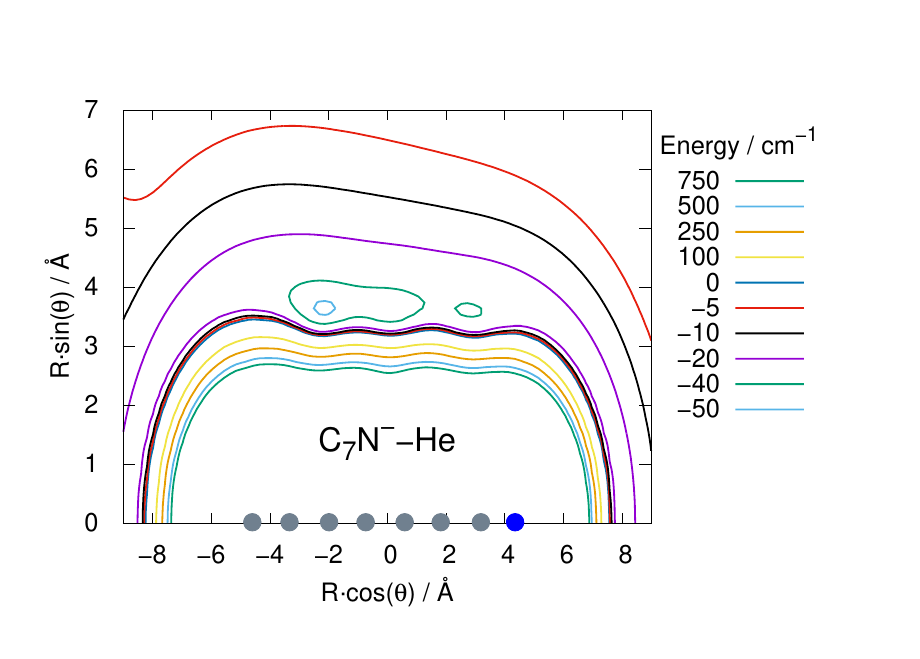}
   \caption{Pictorial representation of the spatial features of the interaction between the C$_7$N$^-$ partner and the He atom. The first C atom is on the left while the end N atom is on the right. See main text for further details.}
        \label{fig: Fig1}
\end{figure}

It is interesting to note from the structure of the isolines that the interaction becomes, at all angles, very repulsive at rather large distances from the c.o.m., a difficult feature to represent, as we have already discussed earlier, and which reflects the unusually long, linear structure of this anionic target. We further see that the attractive regions of its interaction with a neutral He atom are located within a fairly localized valley corresponding to a T-shaped configuration of the van der Waals complex. Finally, the `oscillations` shown by the isolines, as one comes closer to the chain of carbon atoms, indicate the detailed changes of the interaction forces with the approaching He partner as one moves from the carbon atoms to the bonding sections along the molecular chain.

  The multipolar expansion  was carried out to $\lambda$ values up to $\lambda$ = 93, to ensure numerical convergence, employing the procedure discussed more extensively earlier in this Section, where we explain how a larger number of raw points were obtained by adding to the initial selection. The actual features of the multipolar coefficients for the C$_5$N$^-$ partner with the He projectile were discussed in detail in our earlier work reported by ref \cite{BGG23}, hence we shall not be repeating it in the present paper. We only report in Figure 2  a comparison between the lower six coefficients newly obtained for the longer molecular anion, the C$_7$N$^-$, and those we had found earlier for the C$_5$N$^-$ anion, to indicate the relative strength of the angular coupling that will be active for the various radial regions of their interactions with the neutral He atoms. The multipolar coefficients reported in that Figure are usually found to be the more effective during the quantum dynamics couplings of target rotational levels, as further discussed in the next Section.

\begin{figure}[!ht]
   \centering
        \includegraphics[width=0.48\textwidth,angle=+0.0]{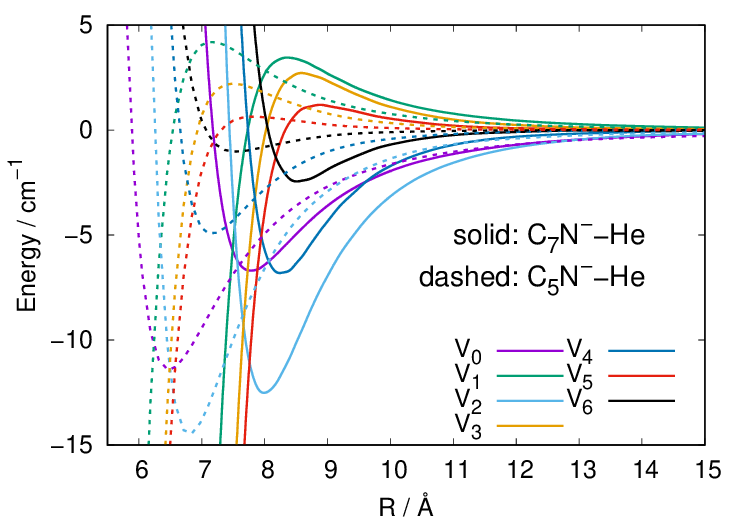}
        \caption{A comparison of the lower radial multipolar coefficients for the interaction potential energy surfaces between the  C$_5$N$^-$ and the C$_7$N$^-$ anions and the He atom. See main text for further comments.}
        \label{fig: Fig2}
\end{figure}

  As it is to be expected, the  interactions with $\lambda$=0 exhibit the largest well depths for their attractive regions, while the next strong interactions come from the $\lambda$=1 and  2 coefficients for both partners. These coefficients also present marked well depths. The significance of having these radial coefficients  providing large anisotropic couplings will be discussed further in the next Section.
  
    The computed rotational constants for the two anions, treated at the  level of spherical, rigid rotors in $^1 \Sigma$ states (see further discussion below) were both  found to be fairly small, as is to be expected for such long molecular chains:the $B_e$ value is 0.046292  cm$^{-1}$ for the C$_5$N$^-$ molecule and 0.01942 cm$^{-1}$ for the C$_7$N$^-$ anion as discussed in \cite{Cernicharo2020} and in \cite{Cern2023}. For a comparison between the two rotational energy ladders which involve the range of states which have been experimentally detected in \cite{Cernicharo2020} and in \cite{Cern2023}, we report in Figure 3 the relative positioning of these states for the two molecules. We clearly see there that the relevant  levels cover in each case  about 20 or so  cm$^{-1}$ and that many more states are available in that range for the longer anion on the right of the figure. Such features indeed suggest the occurrence of strong  coupling between the closely spaced  levels during the dynamical steps, as we  shall further discuss below.

\begin{figure}[!ht]
   \centering
        \includegraphics[width=0.40\textwidth]{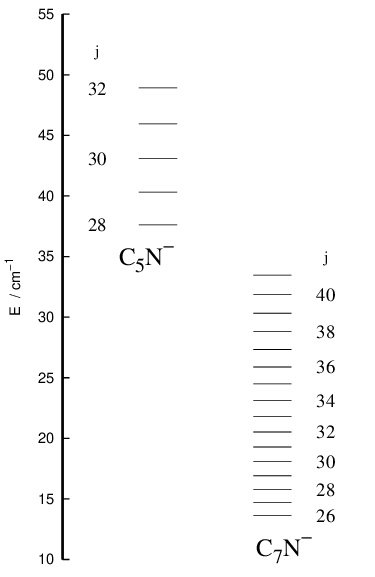}
        \caption{Energy ladders of the rotational levels of the two molecular anions of this study within the range of the observed transitions in the ISM. See main text for further comments}
        \label{fig: Fig3}
\end{figure}

  As a result of the relatively large number of states which can be available over rather small ranges of collision energies, we can argue that at the temperatures reported by the experimental observations of either anion and further discussed below, the reaching of an LTE condition could involve  states up to about $j$ $_{max}$ = 50 or = 60 to be contributing to  rotational state populations up to 50 K. This structural feature will further be discussed later when we shall examine the results from collision and radiative processes  involving both molecular anions.

  \section{ Rotationally Inelastic Quantum Dynamics} \label{sec:dynamics}
  
  Since in our present study no chemical modifications occur during the collision events, the total
scattering wave function can be expanded in terms of asymptotic target rotational eigenfunctions (taken to be spherical harmonics)  whose eigenvalues
are given by $B_ej(j +1)$ where $B_e$ is the first rotational constant mentioned earlier for the present systems \citep{60ArDaxx, 79Secrxx, 97KoHoffxx,79FaGxx}. The channel components for the CC formulation are therefore
expanded into products of total angular momentum eigenfunctions and of radial functions \citep{60ArDaxx}. The latter are in turn the
elements of the solutions matrix which appear in the familiar set of coupled, second order homogeneous differential equations:

\begin{equation}
\left(\frac{d^2}{dR^2} + \mathbf{K}^2 - \mathbf{V} - \frac{\mathbf{l}^2}{R^2} \right) \mathbf{\Psi} = 0
\label{eq.cc}
\end{equation}

where $[\mathbf{K}]_{ij} = \delta_{i,j}2 \mu (E- \epsilon_i)$ are the matrix elements of the diagonal matrix of the asymptotic (squared) wavevectors and 
$[\mathbf{l}]_{ij} = \delta_{i,j}l_i(l_i+1)$ is the matrix representation of the square of the orbital angular momentum
operator. This matrix is block-diagonal with two sub-blocks that contain only even values of  ($l’ + j’$) or only odd values of  
$(l’ + j’)$. 

The  scattering observables are thus obtained in the asymptotic region where the Log-Derivative matrix has a known form in
terms of free-particle solutions and unknown mixing coefficients. For example, in the asymptotic region the solution matrix 
can be written in the form:

\begin{equation}
\mathbf{\Psi} = \mathbf{J}(R) - \mathbf{N}(R) \mathbf{K}
\end{equation}

where $\mathbf{J}(R)$ and $\mathbf{N}(R)$ are matrices of Riccati-Bessel and Riccati-Neumann functions. Therefore, at the end
of the propagation one uses the Log-Derivative matrix to obtain the $\mathbf{K}$ matrix by solving the following linear 
system:

\begin{equation}
(\mathbf{N}' -  \mathbf{Y}\mathbf{N}) = \mathbf{J'} - \mathbf{YJ}
\end{equation}

and from the $\mathbf{K}$ matrix the S-matrix is easily obtained and from it the state-to-state cross sections. We have already
published an algorithm
that modifies the variable phase approach to solve that problem, specifically addressing the latter point and we defer the
interested reader to these references for further details \citep{08LoBoGi,03MaBoGi}.
In the present calculations we first generated the necessary state-to-state
rotationally inelastic cross sections and, once these quantities were known, the required rotationally inelastic rate coefficients $k_{j \rightarrow j'} (T)$  
 were evaluated as the convolution of these cross sections $\sigma_{j \rightarrow j'}$ over a Boltzmann distribution of the  relative translational energy values between partners($E_{\textrm{trans}}$):

\begin{multline}
k_{j \rightarrow j'} (T) =
\left(  \frac{8}{\pi \mu k_{\text{B}}^3 T^3} \right)^{1/2} \\
\times\int_0^{\infty} E_{\textrm{trans}} \sigma_{j \rightarrow j'}(E_{\textrm{trans}}) 
 e^{-E_{\textrm{trans}}/k_{\text{B}} T} dE_{\textrm{trans}} 
\label{eq:rate}
\end{multline}

       The reduced masses  for the C$_5$N$^-$/He and C$_7$N$^-$/He systems were (in units of amu): 3.797372 and 3.8455, respectively. The individual rate coefficients were obtained at intervals of 1 K, starting from 20 K and going up to 50 K. The new scattering calculations for the larger C$_7$N$^-$ anion were carried out using the MOLSCAT suite of codes \citep{MOLSCAT,MOLSCAT2}.
       
      Briefly, We have included 87 rotational states for the C$_5$N$^-$/He system and two rotational states for H$_2$  for our calculations on the C$_5$N$^-$/H$_2$ system . The maximum value of J$_{TOT}$ was 200, depending on the energy. The convergence checks of the results indicated an accuracy of the cross sections around 2-3$\%$. The  spacing of the energy points (in cm$^{-1}$) for the cross section calculations varied from 0.1 up to 15.0, to 5.0 up to 180.0 and then to 25.0 up to the largest energies of about 350.0 cm$^{-1}$.
              In the case of the C$_7$N$^-$/He system we found that, for the studied transitions,to include rotational states up to j$_{max}$=60 were enough, and the maximum value of J$_{TOT}$ was 180, depending on the energy. Regarding the energy points, we did not use a regular grid. To ensure the convergence of the rate constants to the same values mentioned before  in the same interval of temperatures, we have considered, for all initial states, collision energies from 0.001 to 2.5 cm$^{-1}$ and total energies from 10 to 300 cm$^{-1}$. 
              
             We should further note here that at the highest collision energies considered, i.e. above 140 cm$^{-1}$, we have switched from the full Coupled Channel (CC) approach to the reduction scheme of the Coupled States (CS) approach as described in \citep{MOLSCAT,MOLSCAT2}. Furthermore, at some of the selected high energy values we additionally run calculations at the CC level for comparison and found that the actual cross sections remained within 2-3$\%$ of the CS values.

  \section{Einstein Coefficients For Radiative Emission} 
  \label{sec:A_coeffs}
  
  Another important process for characterizing the internal state distribution of the anions  is their interaction
with the surrounding radiative field, also known as black-body radiation. The total emission (em) transition rates from an excited state $X$ can be written as sum of stimulated (sti) and spontaneous (spo) emission rate coefficients,  as discussed in detail in, e.g.: \cite{03BrCaxx} and as defined below:

\begin{equation}
X^{em}_{k \to i} = X^{sti}_{k \to i} + X_{k \to i}^{spo} = A_{k \to i}(1 + \eta_{\gamma}(\nu,T))
\label{eq:tot_emi}
\end{equation}

where $A_{k \to i}$ is the Einstein coefficient for spontaneous emission and 
$\eta_{\gamma}(\nu ,T) = (e^{(h\nu/k_B T)} - 1)^{-1}$ is the Bose-Einstein photon occupation number for the stimulated processes.
In the present study the most important role is played by the spontaneous emission step, since in the fairly low photon densities of the regions of the diffuse molecular clouds, which  we are considering relevant here, make the stimulated processes less likely to occur.

The Einstein coefficient for dipole transitions is given as
\begin{equation}
A_{k \to i} = \frac{2}{3} \frac{e^2\omega^3_{k \to i}}{\epsilon_0 c^3 h} \left| \int \Psi^*_k \cdot \overrightarrow{r} \cdot
\Psi_i \cdot dr^3 \right|^2
\label{eq:dip_tran}
\end{equation}
where $\omega_{k \to i} \approx 2B_0(j_i + 1)$ is the transition's angular frequency, $\overrightarrow{r}$ is the
charge displacement vector and $\Psi_{k/i}$ are the wavefunctions for states $k$ and $i$ respectively. For pure rotational transitions, eq. \ref{eq:dip_tran} simplifies to 
\begin{equation}
A_{k \to i} = \frac{2}{3} \frac{\omega^3_{k \to i}}{\epsilon_0 h} \mu_0^2 \frac{j_k}{2j_k + 1}
\label{eq:dip_tran_rot}
\end{equation}
where $\mu_0$ is the permanent electric dipole moment of the molecule. Both the dipole values and the chosen rotational constants for the present anions were discussed earlier and given in the previous Section on the structural calculations.

The results from the present calculations, which involve the same transitions as in the  emission lines detected in the Interstellar medium  by \cite{Cernicharo2020,Cern2023}, are given by the data in Table 1.

\begin{table}[]
\renewcommand{\arraystretch}{0.9}
\hskip-1.0cm\begin{tabular}{|l|l|l|}
\hline
\textbf{Transition $j$ $\leftarrow$ $j'$}  &  \textbf{(C$_5$N$^-$) } & \textbf{(C$_7$N$^-$) } \\ \hline
27 $\leftarrow$ 28 &         7.26                   &      1.12                            \\ \hline
28 $\leftarrow$ 29 &          8.07                  &      1.24                            \\ \hline
29 $\leftarrow$ 30 &             8.94              &      1.38                             \\ \hline
30 $\leftarrow$ 31 &              9.87             &      1.52                             \\ \hline
31 $\leftarrow$ 32 &              10.9              &      1.67                             \\ \hline
32 $\leftarrow$ 33 &               11.9             &      1.83                             \\ \hline
33 $\leftarrow$ 34 &               13.0           &      2.01                              \\ \hline
34 $\leftarrow$ 35 &             14.2            &       2.19                             \\ \hline
35 $\leftarrow$ 36 &            15.5            &       2.38                              \\ \hline
36 $\leftarrow$ 37 &          16.8               &        2.59                             \\ \hline
37 $\leftarrow$ 38 &            18.2            &          2.81                          \\ \hline
38 $\leftarrow$ 39 &           19.7           &         3.03                              \\ \hline
39 $\leftarrow$ 40 &            21.3            &          3.28                           \\ \hline
40 $\leftarrow$ 41 &            22.9            &          3.53                             \\ \hline
\end{tabular}
\caption{Einstein Emission coefficients ($A_{ij}$) for C$_5$N$^-$ and C$_7$N$^-$ with $\Delta j=-1$. In units of 10$^5$ s$^{-1}$. }
\end{table}

The emission lines for the C$_5$N$^-$ anion were observed in the IRC +10216 at an estimated  temperature of (37 +/- 6) K. For the case of the C$_7$N$^-$ anion the emission lines were observed in either the TMC-1 or the IRC +10216 CSEs at an estimated  temperature of 26(+/-1.8) K. Given the very small spacing between the levels involved, if the two anions were to be considered under local thermal equilibrium, then the C$_5$N$^-$  molecule would have its rotational states with $j$ values from 25 and up to 35 with a fractional population around 25$\%$. For the case of the C$_7$N$^-$ anion, its rotational states with $j$ values up to 41 would  have a fractional population around 20$\%$. We have therefore computed the rate coefficients involving that range of populated internal levels of the two title anions.

\section{Inelastic rate coefficients} 
\label{sec:rate.coeff.}

  Following the numerical evaluation of the relevant inelastic cross sections indicated in the previous Section, we further implemented the integration process of eq.(10) and employed the computed cross sections to generate inelastic rate coefficients over a range of about 50 K, the expected temperatures relevant to the CSE environments where detection of the present radical occurred, as discussed in our Introduction.

    Examples  for the case of the  C$_5$N$^-$ anion are displayed in the Figures reported in the following. Figure 4 shows excitation processes which  start from the  $j$=25 rotational state of this anion and end up in one of the higher rotational  states  involved in the emission lines experimentally detected in \citep{Cernicharo2020}. This example is meant to show how large the rates would be when considering excitation with fairly large $\Delta$ $j$ values beyond the more efficient cases with  $\Delta$ $j$= $+/-$1 discussed by our earlier work in \citep{BGG23}.

\begin{figure}[!ht]
   \centering
        \includegraphics[width=0.45\textwidth,angle=+0.0]{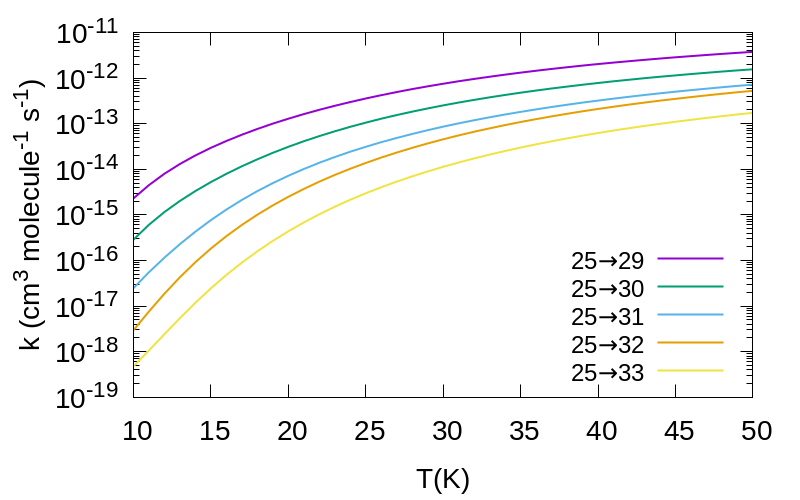}
        \caption{Computed excitation rate coefficients from the $j$=25 rotational state  for the C$_5$N$^-$ for transitions into the observed range of rotational states, over a temperature range of 50 K. See main text for further details.}
        \label{fig: Fig3}
\end{figure}

   Because the excitation processes shown in   Fig. 4 involve large values of the $\Delta$j transition index, we can see that the corresponding rates have fairly small values at the lower temperatures and remain  small up to the highest $T$ values considered. 
   
   On the other hand, if we now consider the excitation processes shown by the data in Figure 5, we clearly see that the rate coefficients  associated with the  $\Delta$j=+1 value are much larger than the ones corresponding to the larger $\Delta$j reported in Figure 4. This is indeed to be expected from the changes in the relative energy gaps  involved in the excitation processes, and as discussed in detail by us in our earlier work in ref \citep{BGG23}. It follows from these findings that the efficient collisional re-populations of the  levels  in the observational emission lines of ref. \cite{Cernicharo2020} would more easily occur from the contiguous levels that are active during those emission processes.

\begin{figure}[!ht]
   \centering
        \includegraphics[width=0.45\textwidth,angle=+0.0]{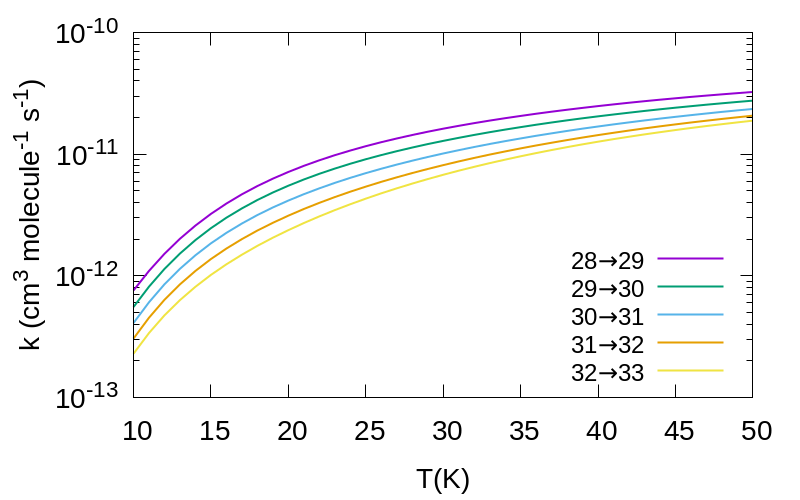}
        \caption{Computed excitation rate coefficients for$\Delta$j=+1 transitions over the observed range of rotational states  for the C$_5$N$^-$. The data cover a temperature range up to 50 K See main text for further details.}
        \label{fig: Fig5}
\end{figure}

Looking at the data in Figure 5 which involve excitations associated  with $\Delta$j=+1 transitions between the levels detected by the experimental work of \cite{Cernicharo2020}, we clearly see  that all the rate coefficients are uniformly much larger than those associated with $\Delta$j values from +4 to +8 reported in Figure 4. These rates are therefore several orders of magnitude larger than those shown by Figure 4.
 
 If we now turn to  the de-excitation ( collisional cooling) processes (presented in  Figure 6), and involving transitions with $\Delta$ $j$=-1, we see that the rate coefficients are all of the order of  10$^{-11}$ cm$^3$molecule$^{-1}$s$^{-1}$, a size which makes them comparable with the corresponding excitation rates discussed in  Figure 5. These will be the quantities that we shall further compare with the competing radiative processes, those driven by the large dipole moments of the present anions.
 
 \begin{figure}[!ht]
   \centering
        \includegraphics[width=0.45\textwidth,angle=+0.0]{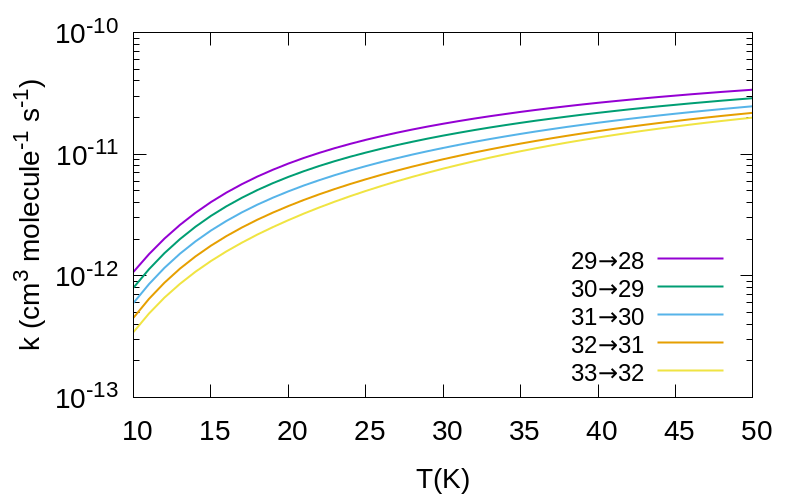}
        \caption{Computed de-excitation (cooling) rate coefficients from a variety of initial   rotational states for  C$_5$N$^-$ the range of temperature values is up to 50 K. See main text for further details.}
        \label{fig: Fig6}
\end{figure}

  The set of calculations for the larger anionic chain, detected just recently by \cite{Cern2023}, and also involving excitation and de-excitation processes for the levels observed in their emission lines (see: \cite{Cern2023}), are given in Figures 7 and 8.

\begin{figure}[!ht]
   \centering
        \includegraphics[width=0.45\textwidth,angle=+0.0]{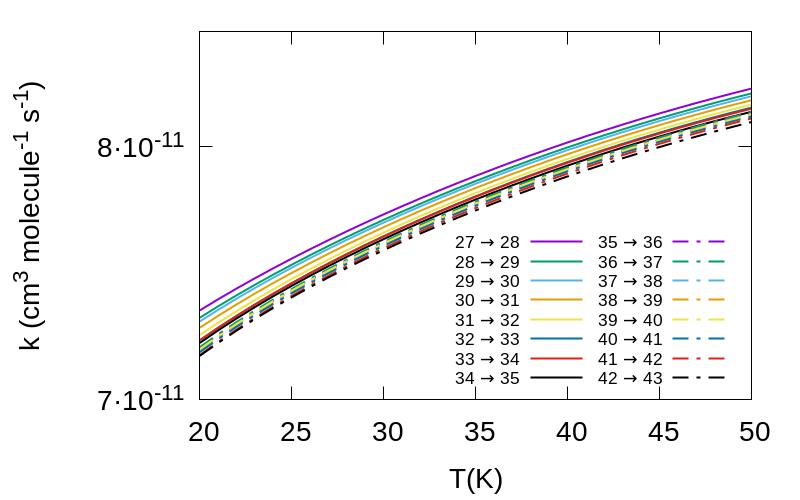}
        \caption{Computed excitation  rate coefficients from a variety of initial   rotational states for  C$_7$N$^-$ in collision with He atoms. The range of temperature values is up to 50 K. See main text for further details.}
        \label{fig: Fig7}
\end{figure}

\begin{figure}[!ht]
   \centering
        \includegraphics[width=0.45\textwidth,angle=+0.0]{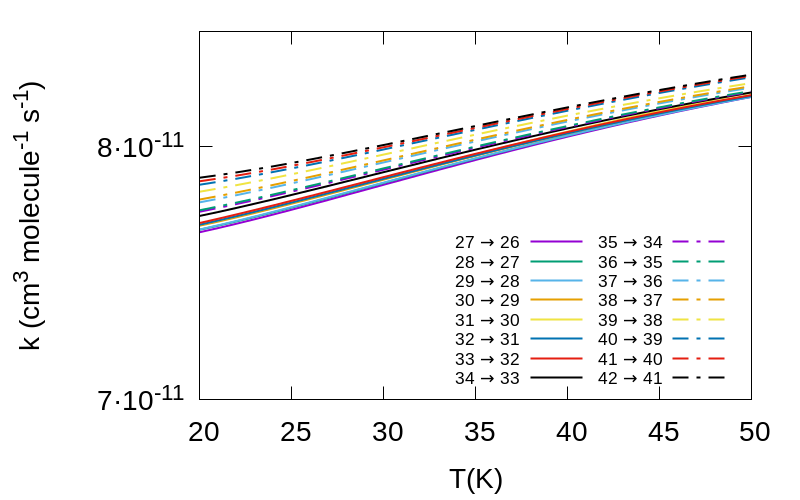}
        \caption{Computed de-excitation (cooling) rate coefficients from a variety of initial   rotational states for  C$_7$N$^-$ in collision with He. The range of temperature values is up to 50 K. See main text for further details.}
        \label{fig: Fig8}
\end{figure}

The computed excitation rate coefficients are now  nearly one order of magnitude larger than those pertaining to the shorter linear anion (see the previous Figure 5). Their range of values indicate how its smaller energy spacings  shown by the Figure 3 are causing the corresponding excitation processes to be more efficient, hence producing larger collision rate coefficients. A similar behaviour can also be gleaned by the results for the de-excitation (cooling) processes, reported in Figure 8. In this case, however, the differences with the results we had found for the C$_5$N$^-$ at lower temperatures turn out to be smaller and therefore the two sets of de-excitation processes are very similar in size over the same range of temperatures.

It is interesting at this point to also compare the inelastic rate coefficients just obtained for the long chains discussed here with those which we have calculated in the past for smaller anions with larger energy spacing between rotational states: the C$_2$H$^-$ and C$_2$N$^-$ interacting with He atoms \citep{20FrGoMa} and the CN$^-$ interacting with He and H$_2$ partners: \citep{20GoGiMa}. In the case of the former anions, in fact, our earlier calculations of the excitation rate coefficients involving the lower rotational states \citep{20FrGoMa} found such rates to be about one order of magnitude larger than those reported here by Figures 5 and 7. Further, in the case of the CN$^-$ anion, our earlier calculations in \citep{20GoGiMa} also showed that the excitation rate coefficients were again about one order of magnitude larger than those reported here for the longer chains. Such findings suggest that the transitions involving the higher rotational states of the present study yield smaller rate coefficients than when the lower-lying rotational states are involved.

In the Introduction Section we have also pointed out that, besides He atoms, the simple hydrogen molecule is also one of the most abundant partner species within the environments where these anions have been observed. Hence, it is also useful to look at the collision-driven processes where the hydrogen molecule is the relevant partner. In our earlier work on the C$_5$N$^-$ we have already considered excitation and de-excitation transitions induced by collision with the H$_2$ partner \citep{BGG23}. Hence, the relevant collision processes among the  experimentally detected levels  are reported in Figure 9. 

 \begin{figure}[!ht]
   \centering
        \includegraphics[width=0.40\textwidth,angle=+0.0]{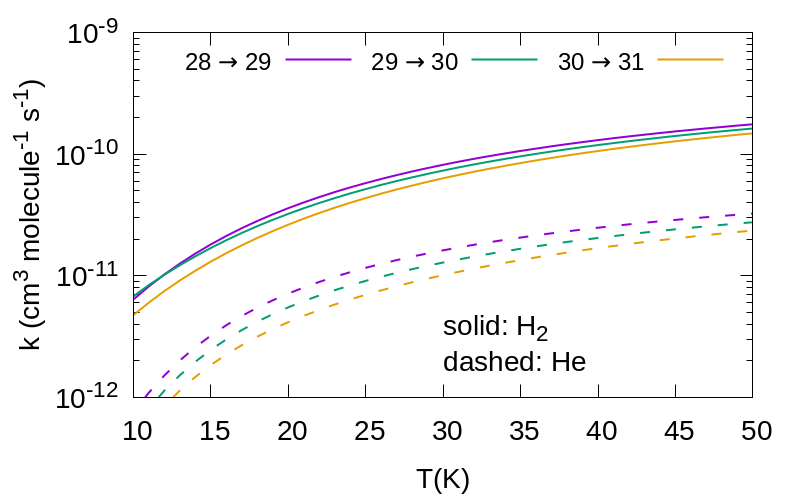}
        \caption{Comparing the computed collision-driven excitation processes into the higher rotational states of C$_5$N$^-$, for the He projectile and the H$_2$(j=0) molecular partner.  See main text for further details.}
        \label{fig: Fig9}
\end{figure}

 \begin{figure}[!ht]
   \centering
        \includegraphics[width=0.40\textwidth,angle=+0.0]{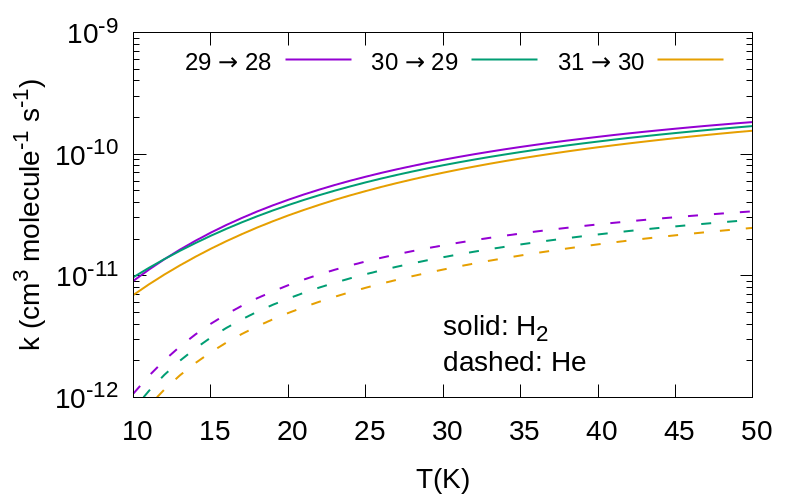}
        \caption{ Same comparison of collision-driven transitions as in the previous Figure 9 but involving here the de-excitation processes in C$_5$N$^-$ induced by both the He  and the H$_2$ projectiles. See main text for further details.}
        \label{fig: Fig10}
\end{figure}

  We clearly see from the comparison reported by Figure 9 that the excitation rate coefficients from collisions with the $p-H_2$ molecule are much larger, by about one order of magnitude on average, than those obtained for the He projectile, thus indicating that both baryonic partners could substantially contribute to a possible, collision-driven, local termal equilibration between the rotational levels of this target anion.

   The de-excitation processes involving  the j=28, 29 and 30 rotational states of the same anion, induced by collision with p-hydrogen molecules, are reported by Figure 10, where comparison is made  with the same transitions induced by collision with He atoms. We see once more that such rates are now about one order of magnitude larger in comparison with those obtained for the He partner. Given the expected abundances of the H$_2$ molecule, we therefore also see that collision-driven population changes could be fairly significant for this partner, as they have already been shown to be important for the case of the He partner.
   
   To acquire a more quantitative understanding of the size differences between the state-changing collision rate coefficients for the two types of partners with the C$_5$N$^-$ anion, we report in Figures 11 and 12  the ratios between the two coefficient values for a sampling of three of the levels involved and over the relevant range of temperatures.

   \begin{figure}[!ht]
   \centering
        \includegraphics[width=0.40\textwidth,angle=+0.0]{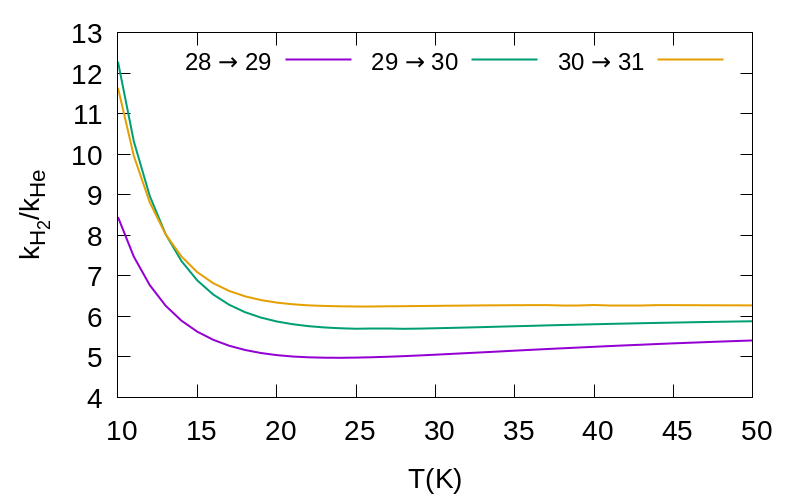}
        \caption{ Computed numerical ratios between the two sets of rate coefficients discussed in this work for the C$_5$N$^-$ anion in collision with H$_2$ and with He. The ratios involve, as examples, three of the relevant rotational states observed experimentally. They are given over a range of temperatures in line with that experimentally attributed to the relevant molecular clouds. See main text for further details.}
        \label{fig: Fig11}
\end{figure}

\begin{figure}[!ht]
   \centering
        \includegraphics[width=0.40\textwidth,angle=+0.0]{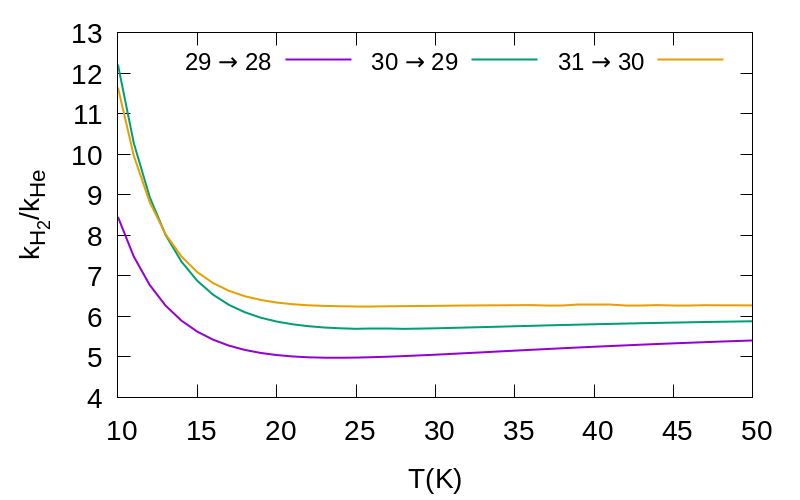}
        \caption{Same ratios as those discussed in Figure 11 but involving instead the de-excitation rate coefficients between the same rotational states. See main text for further details.}
        \label{fig: Fig12}
\end{figure}

It is clear  from the data in those Figures that the rate coefficients pertaining to the H$_2$ projectile are uniformly larger than those obtained from the He partner: their dependence on $T$ at the lowest temperatures above threshold peaks at factors of 10 and larger while becoming fairly constant in the range of $T$ values between  15 K and 50 K,  covering the observed temperatures in the molecular clouds where the title anions have been recently detected: \cite{Cernicharo2020, Cern2023}. This is an interesting result  since, in the case of a linear anion of similar structure, e.g. for the C$_7$N$^-$ discussed in this work, one could conceivably resort to scaling its present data just obtained for the He partner  to estimate the rate coefficients needed for the H$_2$.  This option and its results will be further discussed in the following Section 6.

\section{Analysis of Critical Densities} 
\label{sec:crit_dens}

The assumption of a local thermodynamic equilibrium in different regions  of the interstellar medium is expected, in general,  to hold whenever the population of the excited levels under consideration  is likely to be given by the  Boltzman's law. In the present situation, this  might happen whenever  the rates of spontaneous emission from the internal levels of the polar anions are   smaller than the rates of  state-changing by collision with the most abundant partners present in that ISM region. This implies that  the density in  the interstellar gas for the  partners should be significantly larger than some  critical value  so that the LTE assumption can be considered to be physically viable.
   The definition of a critical density  (e.g. see: \citet{19LMSH,FG21}) is given as follows:

\begin{equation}
n_{\textrm{crit}}^i(T) = \frac{A_{ij}}{\sum_{j \neq i} k_{ij}(T) }
\label{eq:critD}
\end{equation}

where the critical density  for any $i^{th}$ rotational level  is therefore obtained by giving equal weights to the consequences  of
either  the collision-induced or the spontaneous emission processes. We have taken the  rate coefficients discussed in Section
5,  involving the rotational states which have been identified by the detection of our present anions, as discussed in the previous Section. We have also employed the computed spontaneous decay Einstein coefficients discussed  and presented in Section 4. It is also worth mentioning here that there is no sum over $j$ values for the terms at the numerator since the dipole-driven contribution is by far the largest from any given initial $i$th state.

Although the values of the particle densities is considered to vary widely in different ISM environments,so that several general models of the conditions in the molecular clouds suggest variations from the situation of the diffuse clouds (around 10$^6$ cm$^{-3}$) to dense clouds (around 10$^{10}$ - 10$^{12}$ cm$^{-3}$) as discussed, e.g.,  in \citep{06STMC, AC06} and in the references reported there. On the other hand, the specific conditions in the molecular clouds where the present anions have been detected indicate that their densities are expected to be around 10$^4$ - 10$^5$ cm$^{-3}$. Hence, from our computed collision de-excitation rates of, say, Figure 9 for example, we can observe that the above values of expected particle densities can provide a specific range of collisional de-excitation efficiency in units of s$^{-1}$. Hence, in the present molecular  clouds we obtain collision  de-excitation quantities of the order of about 10$^{-5}$ s$^{-1}$.
         
         From the above considerations it follows that in such environments  the collision-induced state-changing processes are faster than the spontaneous radiative emissions reported, for example,  for the C$_5$N$^-$ anion by our calculations in Table 1, while in the more diffuse environments the spontaneous decay channels could be comparable with the collision-induced ones.

 The results  in Figure 13 were obtained using the collision-induced rate coefficients calculated  earlier by us for the C$_5$N$^-$ partner  \citep{BGG23}. The  fairly large values obtained for the critical densities are mainly controlled by the equally  large spontaneous radiative emission coefficients, as reported in the present Table 1, which appear in the numerator of eq. (\ref{eq:critD}).  

\begin{figure}[!ht]
        \includegraphics[width=0.345\textwidth,angle=270]{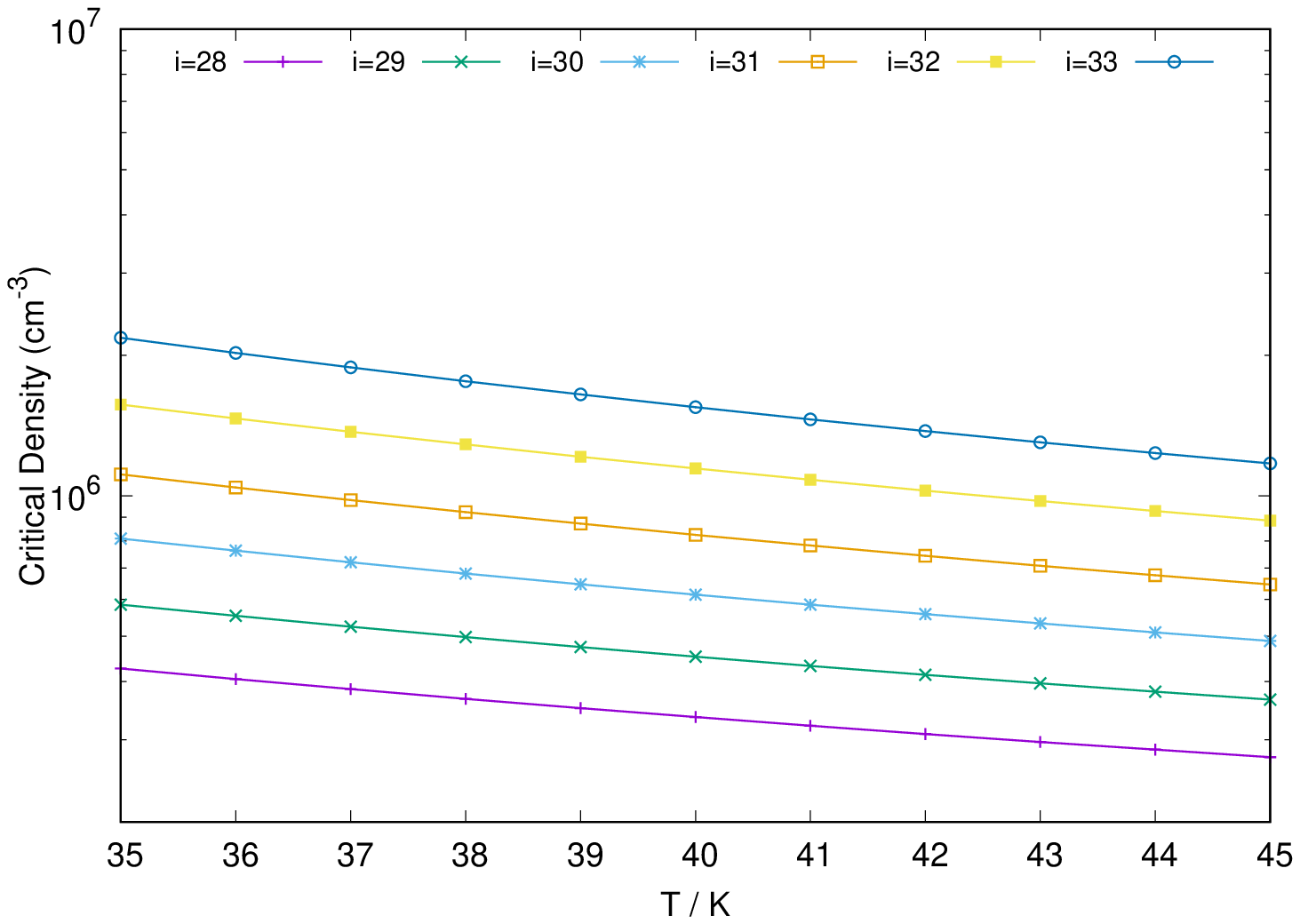}
	\caption{Computed critical densities for the C$_5$N$^-$/He system as obtained from Eq. (\ref{eq:critD}),
        for temperatures from 35 K  to 45 K. Present results are reported for the same rotational levels involved in the detection lines discussed in ref. \citep{Cern2023}.}
        \label{fig:fig13}
\end{figure}
\begin{figure}[!ht]
        \includegraphics[width=0.345\textwidth,angle=270]{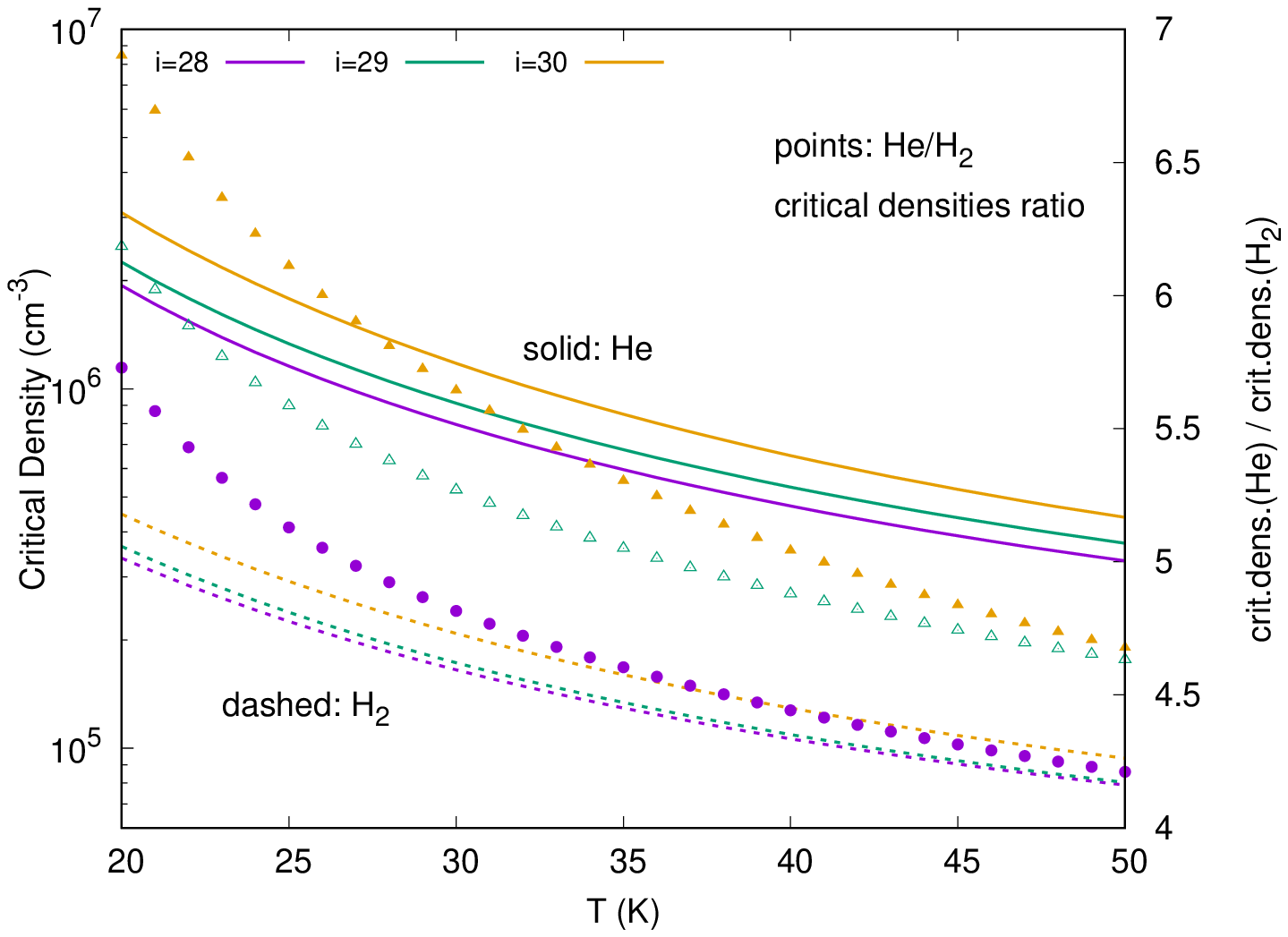}
	\caption{ Comparison between the critical densities obtained for He as a collision partner (solid lines) and those pertaining the molecular hydrogen as the collision partner of the C$_5$N$^-$ anion (dotted lines). The results are reported for some of the rotational levels involved in the detection lines of ref. \citep{Cern2023}.}
        \label{fig:fig14}
\end{figure}
A further example of the relative values of the critical densities for some of the rotational states which have been experimentally detected are reported in Figure 14, where we additionally present  results obtained when  molecular hydrogen is the collision partner of the  C$_5$N$^-$ anion, and compare them with those obtained for the He partner reported in Figure 13. As expected, the larger collision rate coefficients we had found for that molecular partner are now producing smaller values of the critical densities. Within the observed temperature range around 30 K we therefore see that both partners are likely to effectively compete with the radiative processes since the estimated critical densities shown by Figure 14 vary between 10$^5$  $cm^{-3}$ and 10$^6$  $cm^{-3}$. Such range of values  is well within the expected density values of the partner baryonic species in the interstellar environments where these anions have been detected.
We have further extended the previous analysis of the critical densities to the next larger anion recently detected: the C$_7$N$^-$ anion. In that case, we  employ our accurate calculations of the collision rate coefficients obtained for He as a partner atom, while we generate by a scaling procedure those for the H$_2$ molecular partner. Hence, we  use the He/H$_2$ scaling factors we have obtained  for the smaller C$_5$N$^-$ anion and reported by  Figures 11 and 12. That same ratio was thus applied to scale the data for the longer chain anion in collision with He which we have been obtained in this work. The results of the exact and scaled rate coefficients obtained for the longer anion are given in Figures 15 and 16.

\begin{figure}[!ht]
        \includegraphics[width=0.345\textwidth,angle=270]{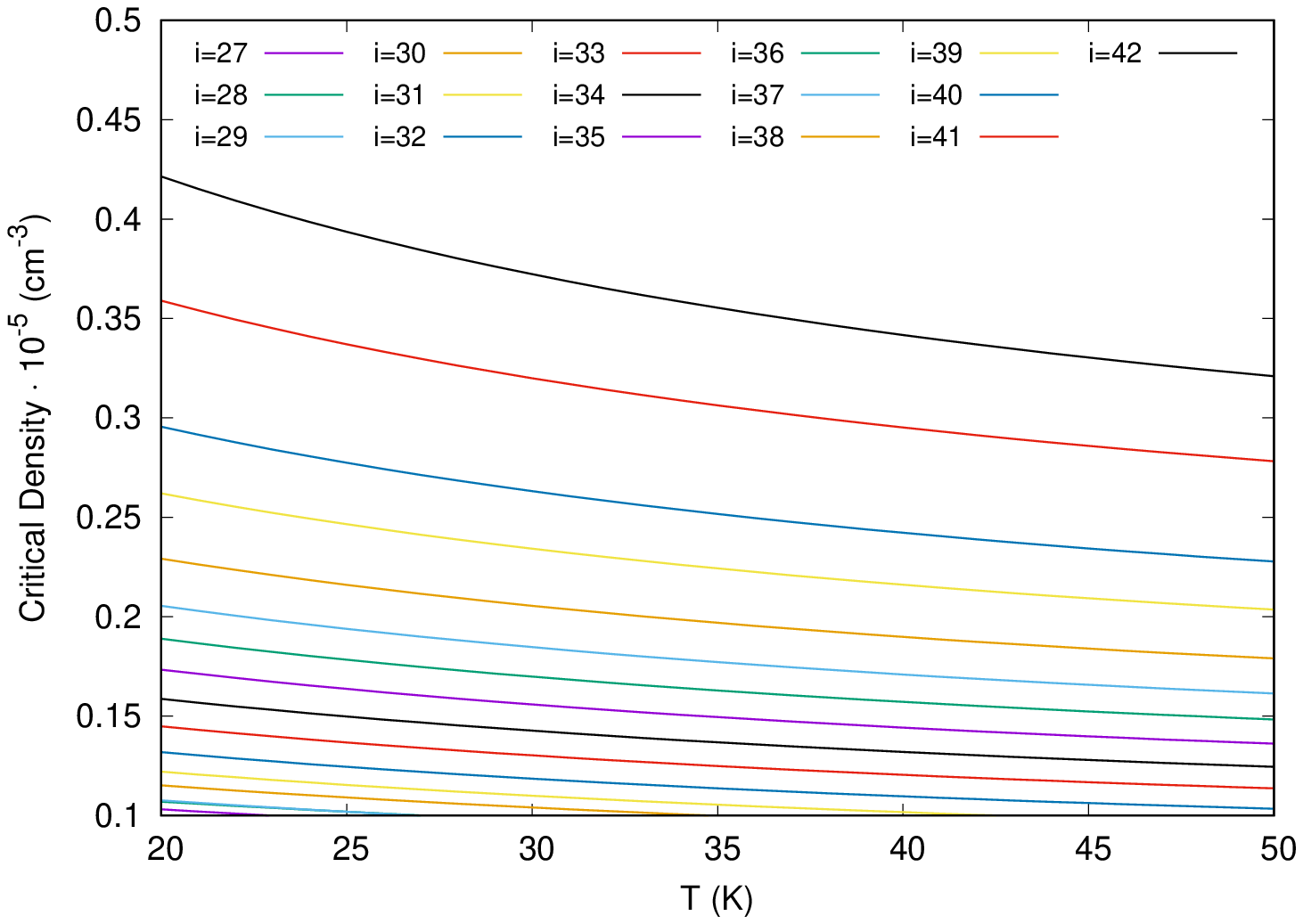}
	\caption{Computed critical densities for the C$_7$N$^-$/He system as obtained from Eq. (\ref{eq:critD}),
        for temperatures from 20 K  to 50 K, using the exact quantum rates discussed in the present paper. The results are reported for the same rotational levels involved in the detection lines discussed in ref. \citep{Cern2023}.}
        \label{fig:fig15}
\end{figure}
\begin{figure}[!ht]
        \includegraphics[width=0.345\textwidth,angle=270]{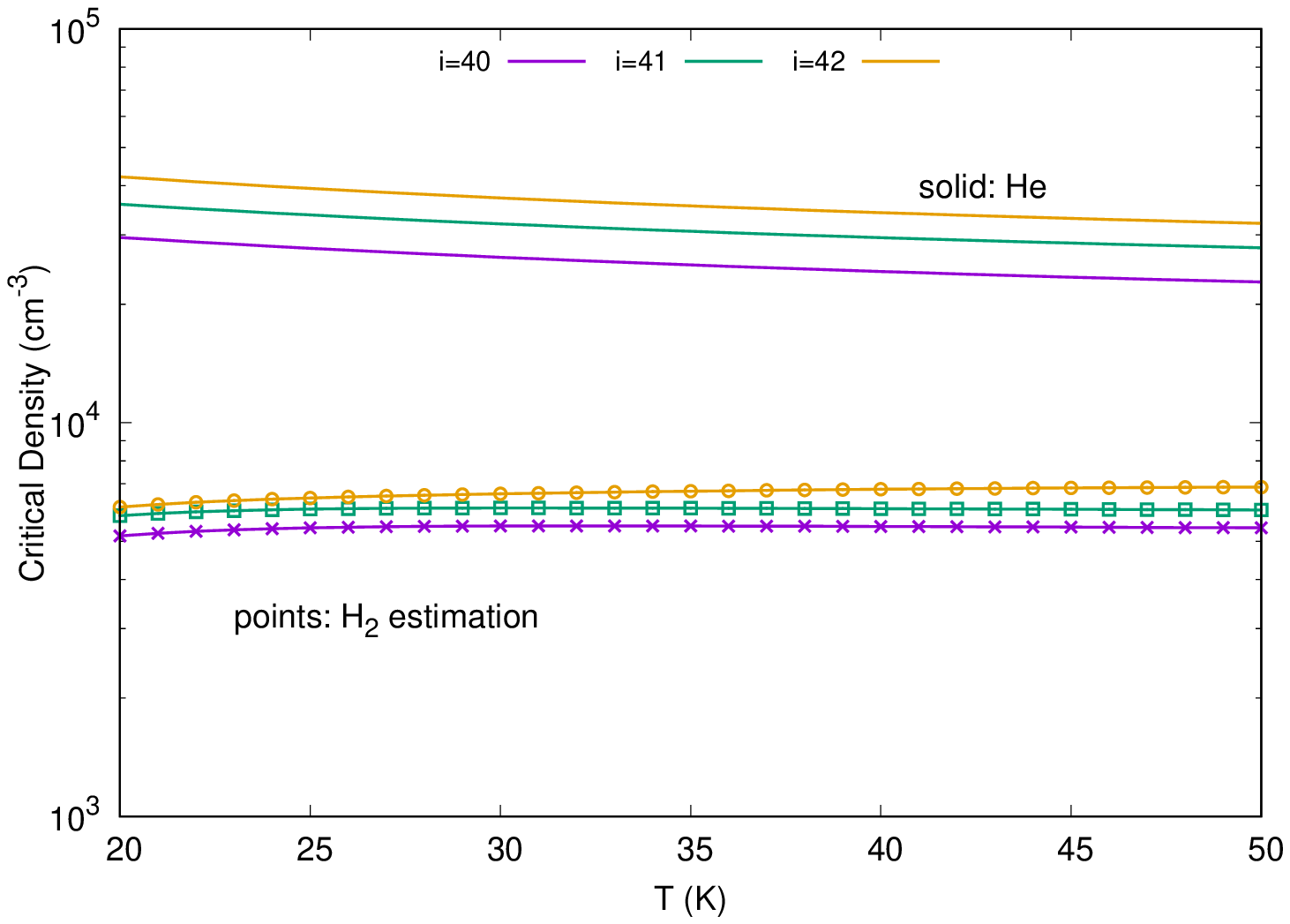}
	\caption{ Comparison between the critical densities obtained for He as a collision partner (solid lines) and those pertaining the molecular hydrogen as the collision partner of the C$_7$N$^-$ anion.The latter rates were obtained from the scaling procedure discussed in the main text. The results are reported for some of the rotational levels involved in the detection lines of ref. \citep{Cern2023}.}
        \label{fig:fig16}
\end{figure}

We clearly see from the data in these two figures that the larger collision rate coefficients, exhibited by the longest cyanopolyyne detected thus far, generate critical densities which are down to vales around 10$^4$cm$^{-3}$ for the He partner and to around 10$^3$cm$^{-3}$  for the H$_2$ molecular partner. Given the baryonic densities around 10$^4$cm$^{-3}$ in the observational environments of these anions, our results suggest that the rotational  state population   
 of the C$_7$N$^-$ anions exhibit critical density values which are large enough to allow the molecules to undergo collision-driven thermalization probabilities which are in competition with their radiative emission de-excitation paths. Under such conditions, therefore, to accurately know the
collision-driven rate coefficients which we have computed  would be important since their knowledge would allow us to establish whether the LTE approximation would  be  valid within their  kinetic networks and whether it would  provide reliable estimates of the relative populations reached under the densities of the CSEs where these anions are present. The knowledge of the actual collision rates with abundant partners like He (and H$_2$) is therefore relevant for  more realistic  modelings of the energy flow processes.

It is interesting to note here that in a recent work by \cite{AMT23} excitation calculations were carried out using  computed collision rate coefficients for a series of cyanopolyynes and polyynes. Their results indicate that the lines from anions accessible to radiotelescopes vary from being subthermally excited to being thermalized as the sizes of the anions increase and link the degree of non-LTE conditions on the H$_2$ volume densities and to the line frequencies. Since the present anions are the largest ones ever observed and not yet discussed in that work, we can argue from our present findings that the critical densities associated to the C$_5$N$^-$ and the C$_7$N$^-$ anions are well within the range of values indicated for the n-H$_2$ column densities that are suggested to vary between 1.0 and 7.5x10$^4$cm$^{-2}$. Hence, it stands to reason to assume that the rotational populations of these longer chains would be more likely to reach thermal conditions in the relevant molecular clouds.

 \section{Present Conclusions} 
 \label{sec:results}

  In the work reported in this paper we have analysed in some detail the behaviour of the excitation and de-excitation (cooling) rate coefficients associated with the rotational states of the two longest cyanopolyyne chain anions, the  C$_5$N$^-$ and the C$_7$N$^-$, which have been  detected so far in  different CSE environments, as we have discussed in the Introduction section. The computed rate coefficients originate from the collisional interactions of these anions with He atoms and H$_2$ molecules, both species well known to be present in significant abundances in the same ISM environments. More specifically, the relative abundance of He with respect to H$_2$ is taken to be about 0.17 \cite{AMT23} and usually considered to have about a 20$\%$ contribution to the state-changing probabilities in comparison with that of the neutral molecular partner. The abundance of the latter is considered to have a local density of the order of 10$^4$  cm$^{-3}$ \cite{AMT23} and therefore relevant for its contribution to the rotational thermalization or sub-thermalization of the emission lines observed for the two title anions.
  
  To further investigate the actual collision rate coefficients for both partners interacting with the two anionic chains, we have employed the accurate \emph{ab initio} evaluation of both He and H$_2$ interactions with the C$_5$N$^-$ already reported by us in our earlier work \cite{BGG23}. For the longer anionic chain of C$_7$N$^-$ we have carried out new \emph{ab initio} calculations of its interaction with He atoms and estimated those pertaining to the H$_2$ molecule by scaling the results for He with the same factors produced for the same partners in the case of C$_5$N$^-$. We have therefore been able to obtain realistic estimates of collision rate coefficients for both anions for the rotational states which had been found to be responsible of their emission lines recently observed in different CSE environments, as described in the Introduction Section.
  
  Using the information we have on the energy spacing between the relevant rotational levels, and with the values we have computed for their permanent dipole moments, we have calculated the Einstein radiative  coefficients acting in the cases of the observed emission lines. The combined use of such coefficients, and of the computed collision rate coefficients involving the same levels, has allowed us to obtain realistic estimates of the critical density values associated with both He and H$_2$ as partners of the title anions. We know, in fact, that  the critical density (and thus the degree of departure from LTE) is very different depending on the dipole moment of the anion and on the frequency of the transition involved. In the present case we have specifically shown that the relevant emission lines which are at the detected temperatures of the cold clouds can be associated, in the case of the C$_5$N$^-$ anion, with critical density values around 10$^6$  cm$^{-3}$ for the He partner at 35 K and around 10$^5$  cm$^{-3}$ for the H$_2$ partner at similar temperatures. The last quantity is indeed comparable with the expected baryonic density of the neutral molecular partner in the molecular clouds where these anions have been observed and therefore suggests that collisional thermalization can be very competitive with the radiative emission rates shown by this anion.
  
  In the case of the next longer molecular cyanopolyyne of our present calculations, we have further found larger collision rate coefficients but comparable radiative coefficients, as shown by our data in Table 1. These findings have produced even smaller critical densities around the same temperatures discussed before: values between 10$^4$ and 10$^5$ cm$^{-3}$ for the case of the He partner and scaled values for the H$_2$ partner of less than 10$^4$ cm$^{-3}$. These results further strengthen the suggestion that collision processes are definitely competitive with the radiative emission processes and can lead to final thermalization of the level populations of these anions in the observed ISM environments. Naturally, the quantities which we have computed in the present study are indeed very important to proceed with a more detailed analysis of the relevant kinetics since, in order to study the abundances and excitations of the present molecular anions in their interstellar sources  we need to further know  the physical parameters of the clouds, like the gas kinetic temperature and the H$_2$ volume density, but also the emission size of anions and the linewidth obtained from observation. In any event, our present, accurate calculations are already able to establish, from the obtained critical densities of both species at the observational temperatures, that collision events are indeed very competitive with the  radiative efficiency indicated by their large dipole moment values and therefore that both long chains should be present with a thermal population of their rotational internal states. 
  
\section{Acknowledgements}

 One of us (L.G-S.)  acknowledges the financial support by Ministerio de Ciencia e Innovación (Spain) MCIN/AEI/10.13039/501100011033 (Ref. PID2020-113147GA-I00 and PID2021-122839NB-I00).   A.V.  acknowledges Grant. No. EDU/1508/2020 (Junta de Castilla y León and European Social Fond). F.A.G. acknowledges the support of the Computing Center of the Innsbruck University, where some of the present calculations were carried out. This research has also made use of the high performance computing resources of the Castilla y León
Supercomputing Center (SCAYLE, www.scayle.es), financed by the European Regional Development Fund (ERDF). We are also grateful to Dr Marcelino Agundez for several discussions and for providing us with a preprint of his recent work on ISM anions.

\section{Supplementary Material}

The following data have been prepared as Supplementary Material: 

(i) original raw points from the $ab$ $initio$ calculations, 

(ii) the multipolar coefficients for the Legendre expansion of the rigid rotor  potential for the C$_7$N$^-$/He system, 

(iii) the computed  inelastic cross sections,  

(iv) the corresponding rate coefficients obtained  from the 2D-RR-PES employed in this work for the C$_7$N$^-$ anion;

(v) The critical density values obtained for both anions and linked to their collision data for either He or H$_2$ are all provided as Supplementary Material. 

            All the above data  are available at  GREDOS: 

\dataset[10.14201/gredos.153158] 
{\doi{10.14201/gredos.153158}.}

\section{ Data Availability}

The data that support the findings of this study are available within the article and in its supplementary material, their accessibility  is described via the link reported in the previous Section.


\bibliography{C5N_C7N}%

\end{document}